\documentclass[epj]{svjour}
\usepackage{graphics}
\begin{document}
\title{The indefinite metric problem revisited and chased away.}
\author
{A. B. van Oosten
}                     
%
%
\institute{Laboratory of Chemical Physics,
University of Groningen,
The Netherlands}
\date{Received: date / Revised version: date}
%
\abstract{
Attempts to quantize light in a manifestly Lorentz covariant manner
fail because of the indefinite metric problem.
Here an error in the interpretation is uncovered
that is at the root of this problem.
} 
\maketitle
\section{Problem}
\label{sec:problem}
The manifestly covariant approach
to the quantisation of electrodynamics
interpretes $A^\mu$ and $A^{\mu+}$
as photon annihilation and creation operators
with commutation relation \cite{gmunu}
\begin{equation}
[A^\mu(\vec {k}),A^{+\nu}(\vec {k}')] = g^{\mu\nu} \delta(\vec {k}-\vec {k}') .
\label{com}
\end{equation}
A single photon state with polarization $\mu$ and momentum $\vec{k}$
is constructed by operating $A^{+\mu}(k)$ on the vacuum state $|0>$.
The in-product of two single photon states is
\begin{eqnarray}
<0|A^\mu(\vec {k}) A^{+\nu}(\vec {k}')|0> &=&
<0|[A^\mu(\vec {k}), A^{+\nu}(\vec {k}')]|0> \nonumber \\
&=& 
g^{\mu\nu} \delta(\vec {k}-\vec {k}')
\label{product}
\end{eqnarray}
For $\mu=\nu=0$ this is a negative definite quantity
which invalidates Eq. \ref{product} as a normalization condition
and is inconsistent with the interpretation of $A^{+\mu}(\vec {k})|0>$
as a single time-like photon state.
In-depth discussions of this problem are given in Refs. \cite{mandl,cohen}. 
Ref. \cite{cohen} even uses a non-covariant approach throughout the text
because of this issue.

\section{Solution}
\label{sec:solution}

First the meaning of the notation $A^\mu$ must be clarified.
It may denote the $\mu$-component of a Minkovski vector,
that is, a number. 
The same symbol also denotes
a Minkovski vector polarized along the $\mu$ direction.

It is an improvement to write $A^{(\mu)}$ for the vector, 
but the notation must also distinguish
a contravariant vector with components $A^{(\mu)\lambda}$
from a covariant one with components $A^{(\mu)}_\lambda$.
The polarization can also be along the contravariant $\mu$-direction,
in which case one should write $A_{(\mu)}^\lambda$ or $A_{(\mu)\lambda}$.

In the proposed notation a single photon state is
either the contravariant $A^{+(\mu)\lambda}(\vec {k})|0>$
or the covariant $A^{+(\mu)}_\lambda(\vec {k})|0>$.
The crucial question is: which bra state belongs to this ket state?
In Eq. \ref{product} covariant bra states
are combined with contravariant ket states,
or vice versa. 
By this choice the infamous minus sign is tacitly introduced.

A {\em positive definite} norm is obtained
by combining the contravariant ket state,
$A^{+(\mu)\lambda}(\vec {k})|0>$,
with the again contravariant bra state $<0|A^(\mu)_\lambda(\vec {k})$,
or, likewise, by combining covariant bra and ket states.
This norm is 
\begin{equation}
<0|A^{(\mu)\lambda}(\vec {k}) A^{+(\nu)\lambda}(\vec {k}')|0>,
\label{norm}
\end{equation}
It takes the covariant form $g^\mu_\nu \delta(\vec {k}-\vec {k}')$,
because $A^{+(\mu)\lambda} = A^+_{(\mu)\lambda}$.

\section{Conclusion}
\label{sec:conclusion}
The indefinite metric problem is merely 
a confusion caused by ambiguous notation.
The covariant quantization procedure is physically sound after all.
It is perhaps surprising
that the fundaments of field theory
are still subject to improvement.


\begin{thebibliography}{}

\bibitem{gmunu}
The convention $g^{\mu\nu}=diag(-1,1,1,1)$ is adopted.

\bibitem{mandl}
F. Mandl and G. Shaw: Quantum Field Theory (Wiley, Chichester, 1984).

\bibitem{cohen}
C. Cohen-Tannoudji, J. Dupont-Roc and G. Grynberg:
Photons and Atoms (Wiley, New York, 1989).

\end{thebibliography}
\end{document}